\documentclass[12pt]{article}

\usepackage{amsmath}
\usepackage{amssymb}
\usepackage{latexsym}
\usepackage{graphics}
\usepackage{psfrag,fancyhdr,epsfig}

\def\beq{\begin{equation}}
\def\eeq{\end{equation}}
\def\bea{\begin{eqnarray}}
\def\eea{\end{eqnarray}}

\begin{document}

\begin{titlepage}

\hfill hep-th/0604182

\vspace*{1.5cm}
\begin{center}
{\bf \Large Brane Models with a Ricci-Coupled Scalar Field }

\bigskip \bigskip \medskip

{\bf  C. Bogdanos, A. Dimitriadis and K. Tamvakis}

\bigskip

{ Physics Department, University of Ioannina\\
Ioannina GR451 10, Greece}

\bigskip \medskip
{\bf Abstract}
\end{center}
We consider the problem of a scalar field, non-minimally coupled to gravity through a $-\xi\phi^{2}R$ term, in
the presence of a Brane. Exact solutions, for a wide range of values of the coupling parameter $\xi$, 
for both $\phi$-dependent 
and $\phi$-independent Brane tension, are derived and their behaviour is studied. In the case of a Randall-Sundrum geometry,
a class of the resulting scalar field solutions exhibits a folded-kink profile.
We go beyond the Randall-Sundrum geometry studying general warp factor solutions in the presence of a kink scalar. 
Analytic and numerical results are provided
for the case of a Brane or for smooth geometries, where the scalar field acts as a 
thick Brane. It is shown that finite geometries with warp factors that asymptotically decrease exponentially are 
realizable for a wide range of parameter values.
 We also study graviton localization in our setup and find that the localizing potential for gravitons with the
 characteristic volcano-like profile develops a local maximum located at the origin for high values of the coupling $\xi$.

\end{titlepage}

\section{Introduction}
The idea of realizing our Universe as a {\textit{defect}} in a higher dimensional spacetime, although not new\cite{RUBSHA},
has received a lot of attention in recent years in the framework of String Theory where {\textit{$D$-Branes}}\cite{DB}, i.e. membranes on
which the fundamental string fields satisfy Dirichlet boundary conditions, play a significant role. In the framework of
String/$M$-theory\cite{MTHEOR} or the AdS/CFT correspondence\cite{ADSCFT}, Brane
models\cite{A}\cite{ADD}\cite{AHDD}\cite{RS1} have revealed new possibilities for
the resolution of the hierarchy problem of particle physics as well as for the relation of gravity to the rest of
fundamental interactions. In $D$-Brane models, Standard Model fields are trapped on the Brane, while gravitons propagate
 in the full higher dimensional space ({\textit{Bulk}}). In an interesting case of a Brane Model with an infinite
 extra dimension, gravitons are localized on the Brane due to the curvature of the extra dimension\cite{RS2}. A solution to
 the Einsten's equations of motion with a flat metric on the Brane and $AdS_5$ geometry in the Bulk exists, provided the
 positive Brane-tension is finely tuned versus a negative Bulk cosmological constant. 
 
 Although the Standard Model fields are assumed to be localized on the Brane, gravity is not necessarily the only field 
 propagating in the Bulk. A number of Brane models with Bulk scalar fields have been constructed\cite{GUB}\cite{DV}, either
 from a theoretical or phenomenological viewpoint\cite{KT1}. Actually, the Brane itself 
 could be a defect substantiated by a Bulk scalar field configuration (a {\textit{``kink"}})\cite{KT}. The presence of a
 Bulk scalar field opens the possibility of a direct coupling of this field to the curvature scalar. A specific form of this
 coupling corresponds to the gravitational term appearing in the so-called tensor-scalar theory of gravity\cite{BD}. A Bulk
 scalar field non-minimally coupled through a coupling of the form $\phi^2\,R$ has also been 
 considered in the Randall-Sundrum framework and numerical solutions have been discussed\cite{FP}.
 
 In the present article we consider a $3$-Brane embedded in $5D$ space endowed with a Bulk scalar field $\phi$, non-minimally
 coupled to gravity through a $-\xi\,\phi^2\,R$ term. We investigate analytically the existence of solutions to the coupled
 system of equations of motion for the metric and the scalar field in the framework of a metric ansatz
 $Diag\left(e^{A(x_5)}\eta_{\mu\nu},\,1\right)$. In the case of the Randall-Sundrum form of the metric we 
 derive analytically a complete set of exact solutions for a range of values of the non-minimal coupling strength $\xi$,
 corresponding to specific choices of the scalar potential. Scalar fields, with or without non-minimal coupling, are often introduced 
 against a given Randall-Sundrum background under the assumption that their effect on the
 background geometry will not be important. 
 We do find exact non-singular scalar field solutions compatible with an exact Randall-Sundrum background, taking into acount 
 the full back-reaction of the field.
 
 We show the existence of a class of solutions for a general warp function with an asymptotic Randall-Sundrum $AdS_5$ 
 behaviour. In all these considerations we allow for a field-dependent Brane-tension. Furthermore, we discuss the existence of 
 smooth $AdS_5$ solutions for which the role of the Brane is played by a ``kink" configuration of the Bulk scalar field itself. Both numerical
 and an approximate analytic treatment of the problem is provided. In particular, we calculate the warp factors for smooth geometries in the
 presence of the kink for different boundary values at the origin and
  obtain various solutions. Although we concentrate on ${\cal{Z}}_2$ symmetric solutions, smooth asymmetric solutions
  are also possible. Through an analytical investigation we verify that for a wide range
  of values of the parameters, we can get warp factors that decrease exponentially and thus provide us 
  with finite geometries. We also 
 find analytical solutions for certain special values of the parameters and different ranges of $\xi$.
  In the final section of this
 paper we study graviton localization in our setup and check the form of the localizing potential for gravitons with the
 characteristic volcano-like profile. We find that for $\xi$ higher than some specific value the potential
  develops a local maximum at the origin, which gradually 
 increases as we move towards higher values of the coupling parameter.
 
\section{The Framework}
Consider a general $5D$ theory of a real scalar field 
coupled to gravity. Allowing only for terms linear in the Ricci scalar, we may
write the general Action as
\beq
{\cal{S}}=\int\,d^5x\,\sqrt{-G}\left\{\,f(\phi)R\,
-\frac{1}{2}(\nabla\phi)^2-V(\phi)-{\cal{L}}_m\,\right\}\,,{\label{Action}}
\eeq
where $f(\phi)$ is, for the moment, a general smooth positive-definite function of the scalar field
$\phi$. $G_{MN}$ is the five-dimensional metric, not to be confused with the Einstein tensor. In the case of a constant $f$, we have the Einstein Action.
 The last term corresponds to $\phi$-independent matter. Note that, the above Action can always be transformed 
 through a {\textit{conformal transformation}} $G_{MN}\rightarrow \tilde{G}_{MN}f(\phi)/2M^3$ into an
 Action where the Ricci scalar enters in the Einstein fashion as $(2M^3)\,R$. Nevertheless, a $\phi$-dependence will
 arise in the matter term giving a theory different than the one we would get in the absence of $f(\phi)$.
 
 The equations of motion resulting from ({\ref{Action}}) are
 \beq
f(\phi)\left(\,R_{MN}-\frac{1}{2}G_{MN}R\,\right)-\nabla_M\nabla_N f(\phi)+G_{MN}\nabla^2f(\phi)=\frac{1}{2}T_{MN}^{(\phi)}+\frac{1}{2}T_{MN}^{(m)}\,,\,
\eeq
\beq
\nabla^2\phi-\frac{dV}{d\phi}+R\frac{df}{d\phi}=0\,,
\eeq
with
\beq
T_{MN}^{(\phi)}=\nabla_M\phi\nabla_N\phi-G_{MN}\left(\,\frac{1}{2}(\nabla\phi)^2+V(\phi)\,\right)\,
\eeq
the energy-momentum tensor of the scalar field $\phi$ and $T_{MN}^{(m)}$ the energy-momentum tensor of (other)
matter.

At this point we shall restrict the metric $G_{MN}$ introducing the {\textit{warped ansatz}}
\beq
G_{MN}=\left(\begin{array}{cc}
e^{A(y)}\eta_{\mu\nu}\,&\,0\\
\,&\,\\
0\,&\,1
\end{array}\right)\,,
\eeq
where $x^M=(x^{\mu},\,x^5)\equiv (x^{\mu},\,y)$ and $\eta_{\mu\nu}$ is the $4D$ Minkowski metric with signature
$(-1,\,1,\,1,\,1)$. We can always choose $A(0)=0$.

The presence of a {\textit{Brane}} introduces an extra term
 \beq
 -\int\,d^5x\,\sqrt{-G}\,\sigma(\phi)\,\delta(y)=-\int\,d^4x\,\sigma(\phi)\,\delta(y)\,\,,
 \eeq
 where the {\textit{Brane tension}} $\sigma(\phi)$ is, in general, $\phi$-dependent. Introducing
  this term in the Action, modifies $T_{MN}$ in Enstein's equations as
  $$\delta T_{MN}=-G_{\mu\nu}\delta_M^{\mu}\delta_N^{\nu}\,\sigma(\phi)\,\delta(y)
  =-\eta_{\mu\nu}\delta_M^{\mu}\delta_N^{\nu}\,\sigma(\phi)\,\delta(y)\,.$$

In what follows we shall ignore the presence of (extra) matter beyond the Bulk scalar field. 
Substituting this metric ansatz into the equations of motion and assuming that the scalar field is just a function of
the fifth coordinate, i.e. $\phi=\phi(y)$, we obtain
\beq 
\frac{3}{2}f(\dot{A})^2+2\dot{A}\dot{f}=\frac{1}{4}(\dot{\phi})^2-\frac{1}{2}V\,,
\eeq

\beq
\frac{3}{2}f\ddot{A}+\frac{3}{2}f(\dot{A})^2+\frac{3}{2}\dot{f}\dot{A}+\ddot{f}=
-\frac{1}{4}(\dot{\phi})^2-\frac{1}{2}V-\frac{1}{2}\sigma(\phi)\,\delta(y)\,,
\eeq

\beq
\ddot{\phi}+2\dot{A}\dot{\phi}-\frac{dV}{d\phi}+\frac{df}{d\phi}\left(-4\ddot{A}-5(\dot{A})^2\right)
-\frac{d\sigma}{d\phi}\delta(y)=0\,.
\eeq
The dot signifies differentiation with respect to the fifth coordinate $y$.

The {\textit{Junction Relations}} at the point $y=0$ where the Brane is located are
\beq
\Delta\dot{\phi}(0)\equiv \dot{\phi}(+0)-\dot{\phi}(-0)=\frac{f\sigma'-\frac{4}{3}\sigma f'}{f+\frac{8}{3}(f')^2}\,,
\eeq
\beq
\Delta\dot{A}(0)\equiv\dot{A}(+0)-\dot{A}(-0)=
-\left(\frac{\frac{1}{3}\sigma+\frac{2}{3}f'\sigma'}{f+\frac{8}{3}(f')^2}\right)\,.
\eeq
The prime signifies differentiation with respect to $\phi$. By $\sigma$ and $\sigma'$ we indicate the values at $\phi(0)$.

Among the above three equations of motion in the {\textit{Bulk}} only two are independent. They can be written as

\beq
V(\phi)=-3f\dot{A}^2-\frac{3}{2}f\ddot{A}-\frac{7}{2}\dot{A}\dot{f}-\ddot{f}\,,\label{V}
\eeq

\beq
\frac{1}{2}\dot{\phi}^2=-\frac{3}{2}f\ddot{A}+\frac{1}{2}\dot{A}\dot{f}-\ddot{f}\,.\label{FI}
\eeq

Let us now restrict the coupling function $f(\phi)$ to be a function quadratic\footnote{Even for a general 
coupling function, we may consider an expansion in even powers of the 
field $f(\phi)\approx f(0)+\frac{1}{2}f'(0)\phi^2+\cdots$ and retain the lowest non-trivial term. Such an expansion would be
valid for small field values ($\phi<<(|2f(0)/f'(0)|)^{1/2}$).} in $\phi$. Introducing a dimensionless parameter
$\xi$ and normalizing it appropriately, we may write
\beq 
f(\phi)=2M^3-\frac{\xi}{2}\phi^2\,.
\eeq
The scale $M$ is related to the $5D$ Newton's constant $\tilde{G}$ as $2M^3=(16\pi{\tilde{G}})^{-1}$. With this choice, we have
$$\dot{f}=-\xi\dot{\phi}\phi\,,\,\,\,\,\,\ddot{f}=-\xi\dot{\phi}^2-\xi\phi\ddot{\phi}$$
and the equations of motion in the Bulk become
\beq
V(\phi)=-\frac{3}{2}\left(2M^3-\frac{\xi}{2}\phi^2\right)\left(\,2\dot{A}^2+\ddot{A}\,\right)+
\frac{7\xi}{2}\dot{A}\dot{\phi}\phi+\xi\ddot{\phi}\phi+\xi\dot{\phi}^2\,,{\label{SP}}
\eeq
\beq
\frac{1}{2}\dot{\phi}^2=-\frac{3}{2}\left(2M^3-\frac{\xi}{2}\phi^2\right)\ddot{A}
-\frac{\xi}{2}\dot{A}\dot{\phi}\phi+\xi\ddot{\phi}\phi+\xi\dot{\phi}^2\,\,.
\eeq
The Junction Relations take the form
\beq
\Delta\dot{\phi}(0)=\frac{\left(2M^3-\frac{\xi}{2}\phi^2(0)\right)\sigma'+\frac{4\xi}{3}\sigma
\phi(0)}{2M^3-\frac{\xi}{2}\left(1-\frac{16}{3}\xi\right)\phi^2(0)}\,,
\eeq
\beq
\Delta\dot{A}(0)=
-\left(\frac{\frac{1}{3}\sigma-\frac{2\xi}{3}\phi(0)\sigma'}{2M^3-\frac{\xi}{2}\left(1-\frac{16}{3}\xi\right)\phi^2(0)}\right)\,.
\eeq
Again, $\sigma$ and $\sigma'$ are the corresponding values at $\phi=\phi(0)$. Note the simplification of the
denominator at the $D=5$ {\textit{conformal value}}\footnote{The conformal value in $D$ dimensions is
$\xi_c^{(D)}=\frac{(D-2)}{4(D-1)}$. See ref.\cite{XD}.} $\xi_c=3/16$.

\section{Randall-Sundrum Metric}

In this section we shall make the definite choice of the warp function $A(y)$ to be the standard Randall-Sundrum
warp function $A(y)=-\kappa |y|$ and impose ${\cal{Z}}_2$ symmetry on the scalar field ($\phi(-y)=\phi(y),\,\,\dot{\phi}(+0)=-\dot{\phi}(-0)$).
Substituting, we obtain the equation for $\phi(y)$ in the $y>0$ Bulk
\beq
\frac{1}{2}\dot{\phi}^2=
\kappa\frac{\xi}{2}\dot{\phi}\phi+\xi\ddot{\phi}\phi+\xi\dot{\phi}^2\,\,.
\eeq
The values on the Brane will have to obbey 
$(1-2\xi)\dot{\phi}^2(0)=\xi\phi(0)\left(\kappa\dot{\phi}(0)+2\ddot{\phi}(0)\right)\phi(0)$. 
Thus, the boundary value $\phi(0)=0$ is possible only with $\xi=1/2$. We proceed distinguishing the two
 cases ($\phi(0)=0$ and
$\phi(0)\neq 0$).

\subsection{Special case with $\phi(0)=0$}

In this case, possible only for $\xi=1/2$, we have the solution
\beq
\phi(y)=\frac{2\dot{\phi}(+0)}{\kappa}\left(1-e^{-\kappa|y|/2}\right)\,{\label{sol.3.1}}.
\eeq
The Junction Relations give
\beq
\sigma=12\kappa M^3\,,\,\,\,\sigma'=2\dot{\phi}(+0)\,.
\eeq
\newpage
Note that this solution is possible only with field-dependent Brane-tension. 
Note also that the first is the standard Randall-Sundrum relation. This special solution has the shape of a {\textit{folded kink}} is plotted in Figure 1. Beyond a small region near the
Brane it reaches a constant value $\phi(\pm\infty)=2\kappa^{-1}\dot{\phi}(+0)$ (Figure 1).
\begin{figure}[!h]
\centering
   \begin{minipage}[c]{0.5\textwidth}
   \centering \includegraphics[width=1\textwidth]{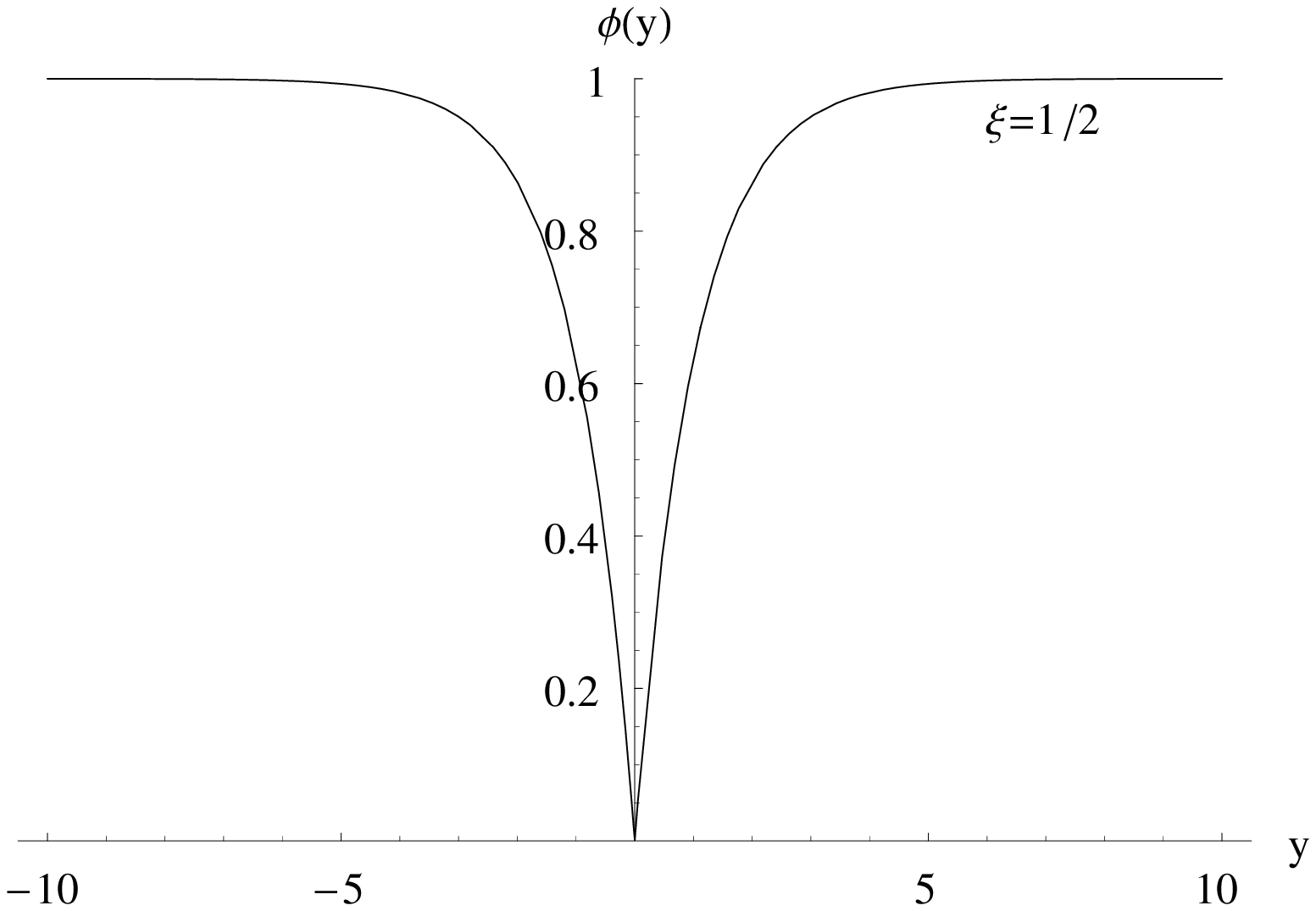}
   \end{minipage}%
   \begin{minipage}[c]{0.5\textwidth}
  \centering  \includegraphics[width=1\textwidth]{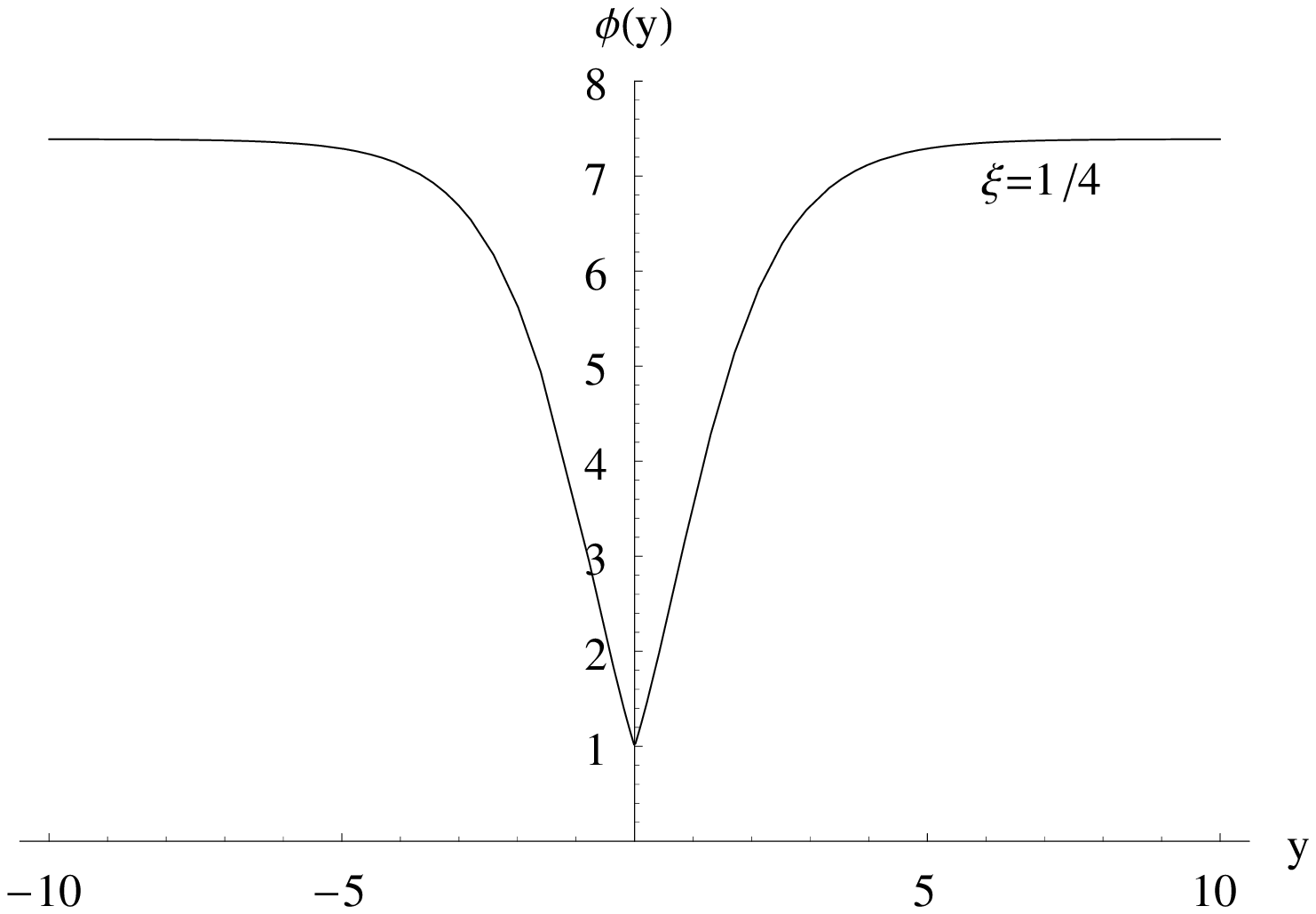}
   \end{minipage}%

   \caption{\textbf{Solutions for a field-independent Brane-tension and different  values of $\bf{\xi}$.}}
   \label{figure1}
\end{figure}%

The required positivity of the coupling function  
\beq
f(\phi)=2M^3-\frac{\xi}{2}\phi^2(y)=2M^3-2\left(\dot{\phi}(0)/\kappa\right)^2\left(1-e^{-\kappa|y|/2}\right)^2>0
\eeq
imposes the boundary value constraint
\beq
(\dot{\phi}(0))^2<\kappa^2M^3\,\Longrightarrow\,(\sigma')^2<\frac{1}{3}\kappa\sigma\,.
\eeq
The scalar potential corresponding to this solution can be obtained to be
\beq
V(\phi)=-2\kappa\sigma-\frac{1}{8}\sigma'^2+\frac{1}{4}(\kappa\phi+\sigma')^2\,.
\eeq
In the expression above, we have made use of the Junction Relations. If we were to start with a general quadratic
potential $V(\phi)=\Lambda+C_1\phi+C_2\phi^2$, the solution ({\ref{sol.3.1}}) and $A(y)=-\kappa|y|$ is possible for
$\Lambda=-2\kappa\sigma+\sigma'^2/8$, $C_1=\kappa\sigma'/2$ and $C_2=\kappa^2/4$.

\subsection{General case}

In the general case $\phi(0)\neq 0$, the equation of motion can be written as
\beq
\frac{1}{2}(1-2\xi)\frac{\dot{\phi}}{\phi}-\kappa\frac{\xi}{2}-\xi\frac{\ddot{\phi}}{\dot{\phi}}=0
\eeq
and leads to the solution

\beq
\phi(y)=\phi(0)\left[1+\alpha
(4-\xi^{-1})
\left(1-e^{-\kappa|y|/2}\right)\,\right]^{\frac{2\xi}{4\xi-1}}\,,{\label{gensol.}}
\eeq

where
\beq
\alpha\equiv \frac{\dot{\phi}(+0)}{\kappa\phi(0)}\,.
\eeq

The Junction Relations take the form
\beq
2\kappa\alpha\phi(0)=\frac{\left(2M^3-\frac{\xi}{2}\phi^2(0)\right)\sigma'+\frac{4\xi}{3}\sigma
\phi(0)}{2M^3-\frac{\xi}{2}\left(1-\frac{16}{3}\xi\right)\phi^2(0)}\,,\,\,\,\,\,
2\kappa=
\frac{\frac{1}{3}\sigma-\frac{2\xi}{3}\phi(0)\sigma'}{2M^3-\frac{\xi}{2}\left(1-\frac{16}{3}\xi\right)\phi^2(0)}\,\,.
\eeq
These two constraints can be rewritten as
\beq
\sigma'=2\kappa\phi(0)(\alpha-4\xi)\,,\,\,\,\,\,
\sigma=6\kappa\left(2M^3-\frac{\xi}{2}\phi^2(0)\right)+4\xi\kappa\alpha\phi^2(0)\,.
\eeq
In order to study whether the positivity of the coupling function and the requirement of a positive
 tension Brane ($\sigma>0$) can be simultaneously satisfied, we consider the four possible
  sign choices of the non-minimal coupling strength parameter $\xi$ and the boundary values parameter $\alpha$. 

They are satisfied\footnote{We choose $\phi(0)>0$.}

1) If $\xi>0,\,\alpha>0$, always.

2) If $\xi<0,\,\alpha<0$, always.

3) If $\xi<0,\,\alpha>0$, only if
\beq
 0<\alpha<\frac{3}{4}+\frac{3M^3}{|\xi|\phi^2(0)}\,.
\eeq

4) If $\xi>0,\,\alpha<0$, only if
\beq
\alpha>\,-\frac{3}{4}+\frac{3M^3}{\xi\phi^2(0)}\,.
\eeq

\subsubsection{Field-independent Brane-tension ($\sigma'=0$)}

To simplify our analysis we may consider separately the case of field-idependent Brane tension ($\sigma'=0$). In this case the Junction Relations simplify to
\beq
6\kappa=
\frac{\sigma}{\left[2M^3-\frac{\xi}{2}\left(1-\frac{16}{3}\xi\right)\phi^2(0)\right]}\,\,
\eeq
and
\beq
\alpha=4\xi\,\,.
\eeq
Notice that the positivity of the Brane-tension is always satisfied, since $2M^3-\xi\phi^2(0)/2+8\xi^2\phi^2(0)/3$ is
positive if the coupling function is positive $(2M^3-\xi\phi^2/2>0$).

The relation $\alpha=4\xi$ simplifies the solution ({\ref{gensol.}}) to
\beq
\phi(y)=\phi(0)\left[1+4
(4\xi-1)
\left(1-e^{-\kappa|y|/2}\right)\,\right]^{\frac{2\xi}{4\xi-1}}\,.
\eeq
For $\xi>\xi_c=3/16$, the quantity in brackets $16(\xi-\xi_c)-16(\xi-\xi_c-1/4)e^{-\kappa|y|/2}$ is positive.

Note that for the special value $\xi=1/2$, if $\phi(y)$ is the solution, so is $\phi(y)+const.$. Also, in the special case 
$\xi=1/4$, the solution takes the form
\beq
\phi(y)=\phi(0)\,e^{2(1-e^{-\kappa|y|/2})}\,.
\eeq
This is shown in Figure 1.

The requirement of the positivity of the coupling function, in the allowed range $\xi_c<\xi$ 
corresponds to the inequality $2M^3\geq \frac{\xi}{2}\phi^2(0)\left[1+4(4\xi-1)\right]^{\frac{4\xi}{4\xi-1}}$. For the special value $\xi=1/4$, this corresponds to $2M^3\geq \frac{e^4}{8}\phi^2(0)$.

For the conformal value $\xi_c=3/16$, the solution reduces to an increasing exponential
\beq
\phi_c(y)=\phi(0)e^{\frac{3}{4}\kappa |y|}\,.
\eeq

For values of the coupling parameter in the range $0<\xi<\xi_c$ the quantity in brackets vanishes at 
$y_0=\pm\frac{2}{\kappa}\ln\left[1+(\xi_c-\xi)^{-1}\right]$, while the exponent is negative, i.e. $\frac{2\xi}{4\xi-1}=-\frac{2\xi}{4(\xi_c-\xi)+\frac{1}{4}}<0$. 
Thus, in this range the solution is singular.

For $\xi$ negative, the solution 
\beq
\phi(y)=\phi(0)\left[\,4(1+4|\xi|)e^{-\kappa|y|/2}-(3+16|\xi|)\,\right]^{\frac{2|\xi|}{4|\xi|+1}}
\eeq
is characterized by an exponent between $0$ and $1$, while the expression in brackets vanishes at
$y_0=\pm\frac{2}{\kappa}\ln\left[1+(3+16|\xi|)^{-1}\right]$. Note that $\dot{\phi}(y_0)=-\infty$. As we shall promptly see, these {\textit{``solutions"}} are not acceptable since the
scalar potential, possesing negative powers of the scalar field, is singular.

The scalar potential corresponding to the solutions found can be immediately obtained from equation ({\ref{SP}}). In order to do that it is usefull to
obtain the derivatives of the solution. They are
\beq
\dot{\phi}(y)=\phi(y)\left(\frac{\xi\kappa}{4\xi-1}\right)
\left[\,\left(1+4(4\xi-1)\right)\left(\frac{\phi(y)}{\phi(0)}\right)^{\frac{1-4\xi}{2\xi}}-1\,\right]
\eeq
and
\beq
\ddot{\phi}(y)=-\dot{\phi}(y)\left(\frac{\xi\kappa}{4\xi-1}\right)
\left[\frac{(2\xi-1)}{2\xi}\left(1+4(4\xi-1)\right)\left(\frac{\phi(y)}{\phi(0)}\right)^{\frac{1-4\xi}{2\xi}}+1\,\right]\,.
\eeq

Substituting the above into ({\ref{SP}}) we obtain
\beq
V(\phi)=-6\kappa^2M^3+\phi^2\left(C_1+C_2\left(\frac{\phi}{\phi(0)}\right)^{\frac{1-4\xi}{2\xi}}+
C_3\left(\frac{\phi}{\phi(0)}\right)^{\frac{1-4\xi}{\xi}}\,\right)\,,
\eeq
where
\beq
C_1=\frac{\xi\kappa^2}{2}\left(3+7\frac{\xi}{4\xi-1}+4\left(\frac{\xi}{4\xi-1}\right)^2\,\right)\,,
\eeq

\beq
C_2=-4\xi\kappa^2\left(1+4(4\xi-1)\right)\left(\frac{\xi}{4\xi-1}\right)\,,\,\,\,\,\,C_3=\frac{C_2^2}{32\xi^2\kappa^2}\,.
\eeq
All powers are positive in the range $0<\xi<\frac{1}{2}$. In the special case $\xi=1/4$, the scalar potential includes logarithmic terms. It is
\beq
V(\phi)=-6\kappa^2M^3+\xi\kappa^2\,\phi^2\,\left[\,-\frac{1}{2}+\frac{3}{4}\ln(\phi/\phi(0))+
\frac{1}{8}\left(\ln(\phi/\phi(0))\,\right)^2\,\,\right]\,\,.
\eeq
For the special value $\xi=1/2$, the scalar potential has the quadratic form
\beq
V(\phi)=-6\kappa^2M^3+\frac{5}{8}\kappa^2\left(3\phi^2-8\phi\phi(0)+25\phi^2(0)\right)\,.
\eeq
For the limiting conformal value $\xi_c=3/16$, all the above coefficients vanish and we obtain a constant potential
$V=-6\kappa^2M^3$. For negative values the appearing powers $\phi^{\frac{1}{2\xi}}$ and $\phi^{\frac{1}{\xi}-2}$ are
negative and, since the solution $\phi(y)$ vanishes at a finite point, the potential is singular.

\subsubsection{General case with field-dependent Brane-tension ($\sigma'\neq 0$)}
In order to investigate the behaviour of the scalar field solution ({\ref{gensol.}}), we first consider the case $\alpha>0$.
 In this case we have a solution increasing near the origin, since $\kappa\alpha=\dot{\phi}(+0)/\phi(0)>0$. The quantity in
brackets is positive, provided $\alpha(4-\xi^{-1})>-(1-e^{-\kappa|y|/2})^{-1}$. 
The lower limit of the right hand side is $-1$, which corresponds to the range $\xi>\frac{1}{4+\alpha^{-1}}$. As examples, consider the cases $\xi=\alpha=1$ and $\xi=\alpha=1/8$. 
The first one corresponds to a positive exponent $2/3$, while the second corresponds to the exponent $-1/2$. They are both 
shown in Figure 2.

\begin{figure}[!h]
\centering
   \begin{minipage}[c]{0.5\textwidth}
   \centering \includegraphics[width=1\textwidth]{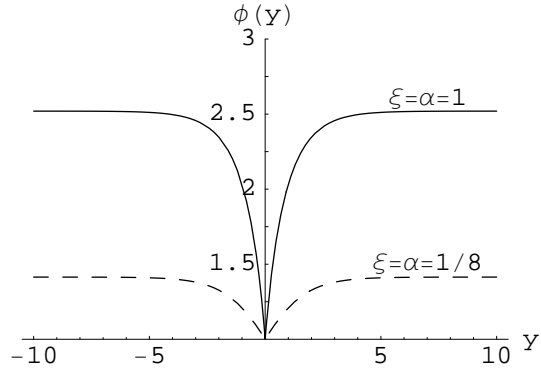}
   \end{minipage}%

   \caption{\textbf{Solution profiles for a field-dependent Brane-tension.}}
   \label{figure2}
\end{figure}%

For values of $\xi$ below this bound there is a point for which the expression in brackets vanishes and, since, the
exponent is negative, there is a singularity. This point is $y_0=-\frac{2}{\kappa}\ln\left[1-\xi/\alpha(1-4\xi)\right]$. 
For the special value $\overline{\xi}=\frac{\alpha}{4\alpha+1}$ the singularity is pushed to infinity and we obtain a purely exponential form for the solution, namely
\beq
\overline{\phi}(y)=\phi(0)\,e^{\kappa\alpha|y|}\,\,.
\eeq

Before we move to consider negative values, let us mention again the special value $\xi=1/4$ which corresponds to
\beq
\phi(y)=\phi(0)\,e^{2\alpha(1-e^{-\kappa|y|/2})}\,\,.
\eeq

As we have remarked earlier, for $\xi>0$ and $\alpha>0$, the positivity of the Brane-tension ($\sigma>0$) is always true.
However, the requirement of a positive coupling function introduces a constraint on the parameters.
 It is sufficient to have $2M^3>\frac{\xi}{2}\phi^2(0)\left[1+\alpha(4-\xi^{-1})\right]^{\frac{4\xi}{4\xi-1}}$. For the special value $\xi=1/4$, this constraint has the form $2M^3>\frac{e^{4\alpha}}{8}\phi^2(0)$.

For negative values $\xi<0$ (and still $\alpha>0$) the solution takes the form
\beq
\phi(y)=\phi(0)\left[1+\alpha(4+|\xi|^{-1})(1-e^{-\kappa|y|/2})\right]^{\frac{2|\xi|}{4|\xi|+1}}\,\,.
\eeq

Note that although the scalar potential has negative powers ($\phi^{-1/2|\xi|}$ and $\phi^{-2-1/|\xi|}$), there is no
singularity, since the scalar field does not vanish anywhere. Note also that for $\xi<0$ and $\alpha>0$ the positivity
 of the Brane-tension introduces a constraint $\alpha<\frac{3}{4}+\frac{3M^3}{|\xi|\phi^2(0)}$.

Let's move now to consider the case $\alpha<0$. Writing the solution as
\beq
\phi(y)=\phi(0)\left[\,1-|\alpha|(4-\xi^{-1})(1-e^{-\kappa|y|/2})\,\right]^{\frac{2\xi}{4\xi-1}}\,,
\eeq
we see that for $\xi>1/4$, the exponent is positive. For $|\alpha|<\frac{1}{4}$ 
the quantity in brackets stays positive. However, for $|\alpha|>\frac{1}{4}$ it is
 necessary to limit the range of $\xi$ to $\xi<\frac{1}{4-\frac{1}{|\alpha|}}$. For the critical value $\overline{\xi}=\frac{|\alpha|}{4|\alpha|-1}$, the solution becomes a decreasing exponential, namely
\beq
\overline{\phi}(y)=\phi(0)\,e^{-\kappa|\alpha||y|}
\eeq
As examples of the solution in the above range, let's consider $\alpha=-1/8,\,\xi=1$ and $\alpha=-1/2,\,\xi=1/3$, shown in Figure 3

\begin{figure}[!h]
\centering
   \begin{minipage}[c]{0.5\textwidth}
   \centering \includegraphics[width=1\textwidth]{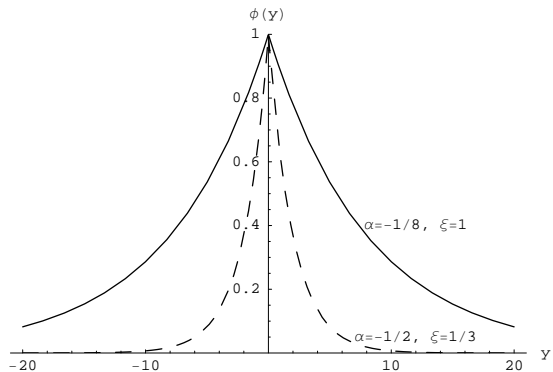}
   \end{minipage}%

   \caption{\textbf{Field-dependent Brane-tension solutions with negative $\alpha$.}}
   \label{figure3}
\end{figure}%

For $|\alpha|>1/4$ and $\xi$ outside of the above range, i.e. $\xi>\frac{1}{4-\frac{1}{|\alpha|}}$, we obtain solutions that vanish at a finite distance from the Brane, namely
 $|y_0|=-\frac{2}{\kappa}\ln\left[1-\xi/|\alpha|(4\xi-1)\right]$. This, again, corresponds to a singular scalar potential due to the negative power $\phi^{1/\xi-2}$ that
appears in it.

For values below $1/4$ but positive, the solution takes the form
\beq
\phi(y)=\phi(0)\left[1+|\alpha|(\xi^{-1}-4)(1-e^{-\kappa|y|/2})\,\right]^{-\frac{2\xi}{1-4\xi}}
\eeq
and gives a smooth decreasing profile, just as seen above.

Before we move to consider negative values of $\xi$, let's consider the special case $\xi=1/4$. In this case, we have
\beq
\phi(y)=\phi(0)\,e^{-2|\alpha|(1-e^{-\kappa |y|/2})}
\eeq
shown for $\alpha=-1$ in Figure 4

\begin{figure}[!h]
\centering
   \begin{minipage}[c]{0.5\textwidth}
   \centering \includegraphics[width=1\textwidth]{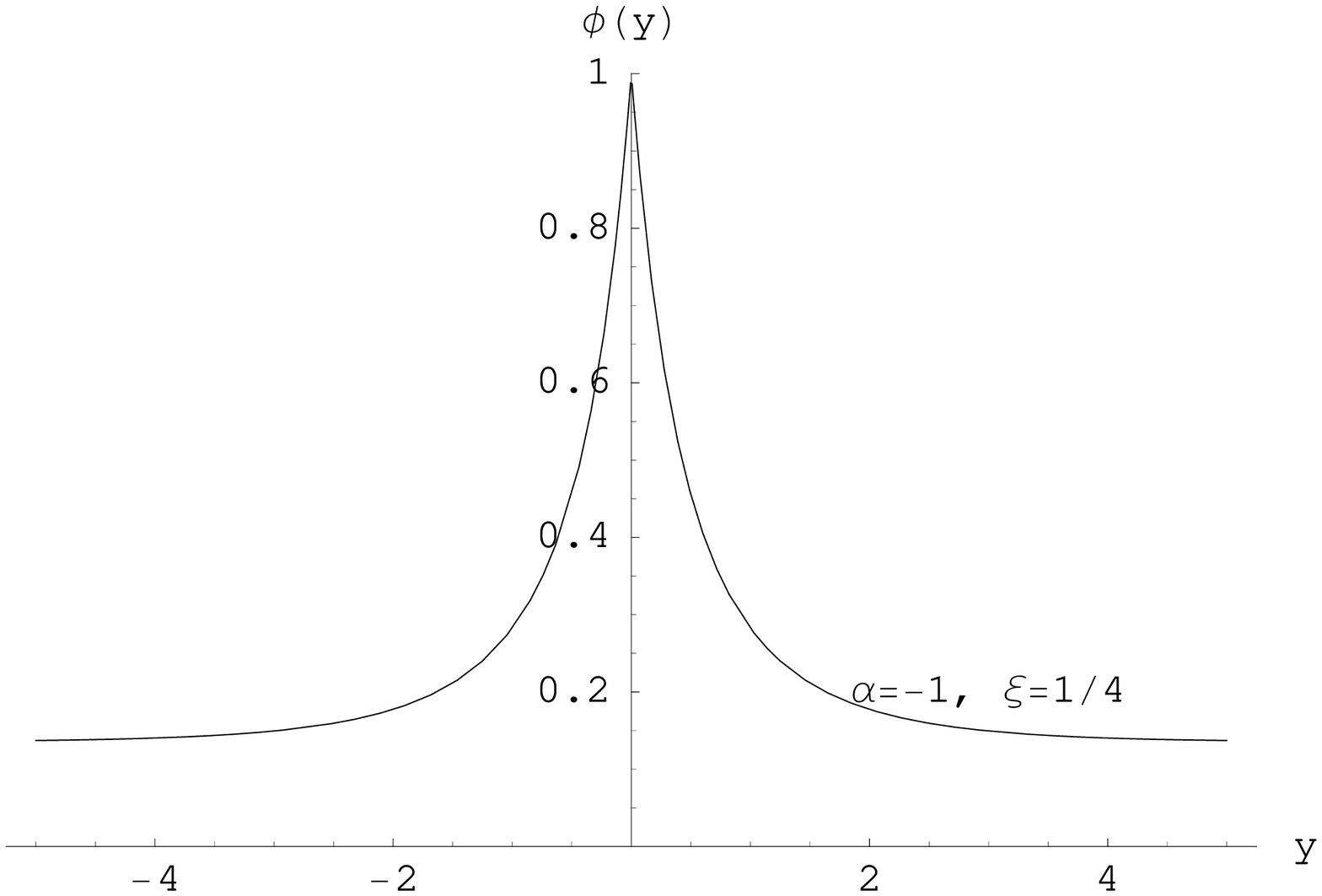}
   \end{minipage}%

   \caption{\textbf{Profile of the solution for the special value $\bf{\xi=\frac{1}{4}}$.}}
   \label{figure4}
\end{figure}%

The positivity of the Brane-tension, for $\xi>0$ and $\alpha<0$, as was found earlier, introduces the constraint
\beq
-\alpha>-\frac{3}{4}+\frac{3M^3}{\xi\phi^2(0)}\,.
\eeq
On the other hand, the positivity of the coupling function, since $\phi(y)$ is a decreasing function, is covered by
$2M^3>\frac{\xi}{2}\phi^2(0)$.

For negative values $\xi<0$, the solution can be written as
\beq
\phi(y)=\phi(0)=\left[\,1-|\alpha|(|\xi|^{-1}+4)(1-e^{-\kappa|y|/2})\,\right]^{\frac{2|\xi|}{4|\xi|+1}}\,.
\eeq
It is easy to see that for $|\alpha|>1/4$, the scalar field vanishes at a finite distance from the Brane, namely 
$|y_0|=-\frac{2}{\kappa}\ln\left[1-|\xi|/|\alpha|(4|\xi|+1)\right]$. This, again, amounts to a singular scalar potential due to the negative powers $\phi^{-1/|\xi|}$ and $\phi^{-(1/|\xi|+2)}$.

The scalar potential has exactly the same form as in the $\sigma'=0$ case ($\alpha=4\xi$), the only difference being a
slight change in the coefficients $C_2$ and $C_3$ which become
\beq
C_2=-4\xi\kappa^2\left(1+\frac{\alpha}{\xi}(4\xi-1)\right)\left(\frac{\xi}{4\xi-1}\right)\,,
\,\,\,\,\,C_3=\frac{C_2^2}{32\xi^2\kappa^2}\,.
\eeq

\section{Beyond Randall-Sundrum}

Let us consider again our original set of the two independent equations (\ref{V}), (\ref{FI}) for the specific choice
 of coupling function $f(\phi)=2M^3-\xi\phi^2/2$. If we do not impose any restriction on the scalar
potential function $V(\phi)$, we can consider the first equation as an equation that {\textit{determines}} the scalar
potential in terms of the functions $\phi(y)$ and $A(y)$. Concentrating on the second equation, we can view it as an
equation for the warp factor, giving a different $A(y)$ for every different {\textit{choice}} of $\phi(y)$ configuration.
Motivated by the form of the solutions found in the Randall-Sundrum case, we may start by introducing a scalar field
configuration in the form of a {\textit{folded kink}} with ${\cal{Z}}_2$ symmetry\footnote{$\dot{A}(+0)=-\dot{A}(-0),\,\,
\dot{\phi}(+0)=-\dot{\phi}(-0)$.}
\beq
\phi(y)=\phi_0\,\tanh(a|y|)\,,\label{KINK}
\eeq
where $\phi_0\equiv a^{-1}\dot{\phi}(+0)$. For the positivity of the coupling function it would be sufficient to require $\dot{\phi}^2(0)<4M^3a^2/\xi$.

Substituting (\ref{KINK}) into the Junction Relations, we obtain, since $\phi(0)=0$,
\beq
\dot{A}(+0)=-\frac{\sigma}{12M^3}\,,\,\,\,\,\,\dot{\phi}(+0)=\frac{\sigma'}{2}\,.
\eeq
Note that the positivity of the Brane-tension requires $\dot{A}(+0)<0$.

In the $y>0$ bulk, we have $\dot{\phi}=a\phi_0(\,1-\phi^2/\phi_0^2)$, $\ddot{\phi}=-2a^2\phi(\,1-\phi^2/\phi_0^2)$. 
Substituting these into the equation of motion, we can write it in the form
\beq
\frac{3}{2}\phi_0\left(2M^3-\frac{\xi}{2}\phi^2\right)X'(\phi)+\frac{\xi}{2}\phi_0\phi
X(\phi)+2a\xi\phi^2+a\phi_0^2\left(\frac{1}{2}-\xi\right)\left(1-\frac{\phi^2}{\phi_0^2}\right)=0\,,
\eeq
where $X\equiv\dot{A}(y)$ and $\ddot{A}=\dot{X}=\dot{\phi}X'(\phi)$. This differential equation can be integrated to give
\beq
X(\phi)=C_0\left(1-\frac{\xi\phi^2}{4M^3}\right)^{1/3}+\phi\,\left\{\,C_1\left(1-\frac{\xi\phi^2}{4M^3}\right)^{1/3}
\,{\,}_2F_1(1/2,1/3,3/2,\,\xi\phi^2/4M^3)\,+C_2\,\right\}\,,
\label{X}
\eeq
where $C_1$ and $C_2$ are given by
\beq
C_1=-\frac{a}{\phi_0}
\left[\,\xi^{-1}-6+
\frac{\phi_0^2}{12M^3}(2\xi-1)\,\right]\,,\,\,\,\,\,\,C_2=\frac{a}{\phi_0}
\left[\,\xi^{-1}-6+\frac{\phi_0^2}{4M^3}(2\xi-1)\,\right]
\eeq
and the integration constant $C_0=\dot{A}(+0)=-\sigma/12M^3$ should be negative in order to have a positive
Brane-tension. 

The metric warp factor will be given by $e^{A(y)}=\exp\left[\,\int\,dy\,X(\phi)\,\right]$. Near the Brane, i.e. for $y\rightarrow 0$ or $\phi\rightarrow 0$, we have
\beq
e^{A(y)}\approx \exp\left[\,-\frac{\sigma}{12M^3}y\,+\,O(y^2)\,\right]\,,
\eeq
which is a pure Randall-Sundrum behaviour.

In the asymptotic region ( $y\rightarrow \infty\,\,\Longrightarrow\,\,\,\phi\sim\phi_0=a^{-1}\dot{\phi}(0)$ ), we have
$$
e^{A(y)}= \exp\left[\,(a\phi_0)^{-1}\int\,\frac{d\phi}{\left(1-\phi^2/\phi_0^2\right)}\,X(\phi)\,\right]\,\approx
$$
\beq 
\,
\exp\left[\,\frac{1}{2a\phi_0}\int\,\frac{d\phi}{\left(1-\phi/\phi_0\right)}\left(\,X(\phi_0)\,+\,(\phi-\phi_0)X'(\phi_0)\,
+\,\cdots\right)\right]\,
\approx 
e^{-\overline{\kappa}\,y\,}\,g(y)\,\,,{\label{ASYMPT}}
 \eeq
 with $\overline{\kappa}\equiv\,-X(\phi_0)$ and
 \beq 
 g(y)\equiv 2^{-\frac{1}{2a}X(\phi_0)}\,
 \exp\left[-\frac{\phi_0}{2a}X'(\phi_0)+\frac{1}{2a}\left(\,X(\phi_0)+2\phi_0X'(\phi_0)\,\right)\,e^{-2ay}+\cdots\right]\,.
 \eeq
 Thus, in the asymptotic region we have a behaviour exponentially close to a Randall-Sundrum behaviour, 
 provided the parameter $\overline{\kappa}$ is positive. This parameter is
 \beq
\overline{\kappa}=\frac{\sigma}{2M^3}\left(1-\frac{\xi\phi_0^2}{4M^3}\right)^{1/3}\,\,-a\,f(\xi,\,\phi_0^2/M^3)\,,
\eeq
where we have introduced a function $f(\xi,\,x)$ defined by
 $$f(\xi,\,x)\equiv
\left(6-\xi^{-1}-\frac{x^2}{12}(2\xi-1)\right)
\left(1-x^2/4\right)^{1/3}{}_2F_1(1/2,\,1/3,\,3/2,\,\xi x^2/4)\,$$
\beq\,\,+\,\,\,\xi^{-1}-6\,+\,
\frac{x^2}{4}(2\xi-1)\,\,\,.
\eeq
The coupling function is positive if $\xi\phi_0^2<4M^3$. The positivity of $\overline{\kappa}$ can always be satisfied with a large
enough Brane-tension parameter $\sigma$. Nevertheless, we can be more concrete by making a choice of $\dot{\phi}(+0)$. We
can take $\phi_0^2=M^3$. Then, the positivity of the coupling function restricts the values of $\xi$ to $\xi<4$. 
The function $f(\xi,1)$ is negative for negative for all allowed values of $\xi$. Thus, the warp factor will always be decreasing. 
The plot of this function of $\xi$ is shown in Figure 5.

 \begin{figure}[!h]
\centering
   \begin{minipage}[c]{0.5\textwidth}
   \centering \includegraphics[width=1\textwidth]{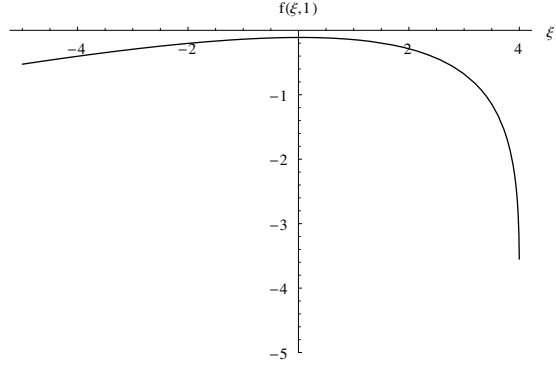}
   \end{minipage}%

   \caption{\textbf{A graph of the function $\bf{f(\xi,1)}$ plotted for all values of $\xi$. THe function maintains a
   negative value. }}
   \label{figure5}
\end{figure}%

We can construct numerical solutions for the function $A(y)$ and study the profile of the warp factor $e^{A\left( y \right)}$
for different values of the parameters of the model. In this case, the brane tension determines the value of the derivative
of $A(y)$ at $y=0$, so it fixes one of the two initial conditions needed for the numerical evaluation. The derivative of the brane
tension with respect to the field, $\sigma'$, is proportional to $a\phi_{0}$. Thus, changing it corresponds
to a new value for $\dot{\phi}(0)$. Note that, although we have restricted ourselves on ${\cal{Z}}_2$-symmetric
solutions, asymmetric solutions are also possible. The
profile of characteristic solutions is plotted in Figure 6.

 \begin{figure}[!h]
\centering
   \begin{minipage}[c]{0.5\textwidth}
   \centering \includegraphics[width=1\textwidth]{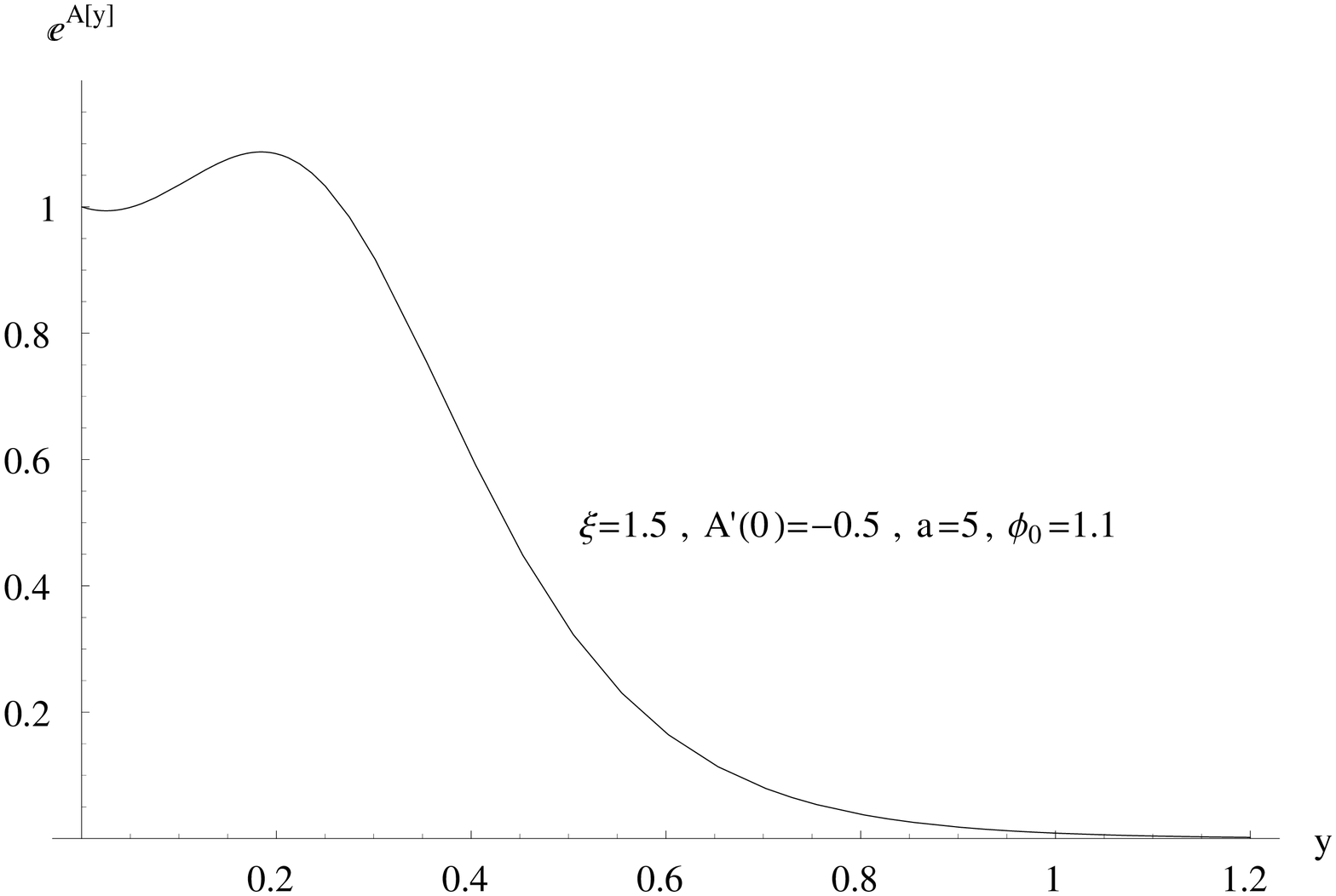}
   \end{minipage}%
   \begin{minipage}[c]{0.5\textwidth}
  \centering  \includegraphics[width=1\textwidth]{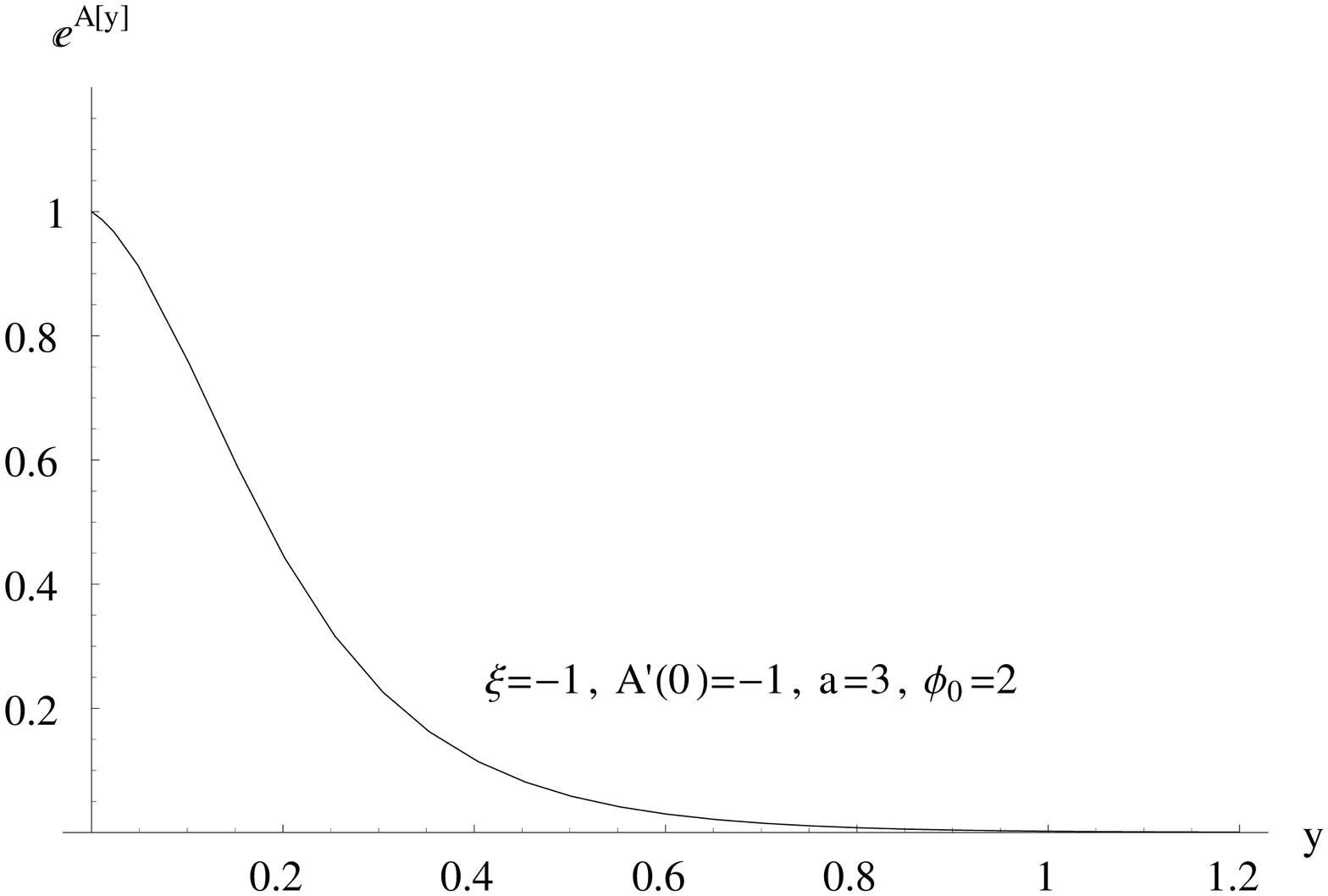}
   \end{minipage}%

   \caption{\textbf{The warp factor for different choices of $A'(0)$, $a$, $\xi$ and $\phi_0$ in units of $2M^3$.} }
   \label{figure6}
\end{figure}%

 We see that, in general, the warp factor resembles the Randall-Sundrum decreasing
exponential. Yet, for a range of values in the parameter space we get solutions which deviate slightly from this form. As
the brane tension becomes smaller and $\xi$ takes higher positive values, the warp factor exhibits a peak close to the
Brane, before it starts decreasing again. This peak is amplified as we approach the value of $\xi$ for which the
coupling function $f(\phi)$ tends to zero.

\section{Smooth Spaces}
As we saw in the last example, the presence of the Brane was not essential to obtain a localized warp factor. In this
section we shall consider solutions $A(y)$ when the {\textit{``Brane"}} is the scalar field configuration itself. Such an
example is well known in the $\xi=0$ case. Introducing a standard kink ($\phi=\phi_0\tanh(ay)$) into the $\xi=0$ 
equation of motion, we obtain in the ${\cal{Z}}_2$ symmetric case ($\dot{A}(0)=0$)
\beq
e^{A(y)}=\left(\cosh(ay)\right)^{-\gamma}e^{-\frac{\gamma}{4}\tanh^2(ay)}\,, 
\eeq
with $\gamma=\phi_0^2/9M^3$.

Smooth solutions of the Bulk equations of motion are also present for $\xi\neq 0$. It is not
 difficult to see that the metric choice 
\beq
e^{A(y)}=\left(\cosh(ay)\right)^{-\gamma}\,,
\eeq
corresponds to the same scalar field solution $\phi(y)=\phi_0\,\tanh(ay)$ with
\beq
\gamma=2\left( \xi^{-1}-6\right)\,,\,\,\,\,\phi_0=a^{-1}\dot{\phi}(0)=(2M^3)^{1/2}\sqrt{\frac{6(1-6\xi)}{\xi(1-2\xi)}}  \,.
\eeq
This solution exists for $0<\xi<1/6$ and only for the above specially chosen value of $\phi_0$. The curvature scalar
of this space is $R=4a^2\gamma\left(1-(1+5\gamma/4)\frac{\phi^2}{\phi_0^2}\right)$. Note the asymptotic $AdS$ value
$R(\infty)=-20 a^2(\xi^{-1}-6)^2$. The scalar potential corresponding to this solution is a quartic function of the
scalar field with tuned $\xi$-dependent coefficients. Note that this solution is a particular case of ({\ref{X}}). The above
choice of $\phi_0$, together with the choice $C_0=0$, corresponds to
\beq
C_1=0\,,\,\,C_2=-\frac{2a}{\sqrt{12M^3}}\sqrt{\xi^{-1}(1-2\xi)(1-6\xi)}\,.
\eeq       

It is interesting that the same metric choice corresponds also to the solution 
\beq
\phi(y)=\phi(0)\left(\cosh(ay)\right)^{-1}
\eeq
with $\phi^2(0)=12M^3(\xi^{-1}-6)/(3-16\xi)$ defined in the same $\xi$-range.

Another intersting solution for the metric is defined by the choice $\phi_0=2\sqrt{\xi^{-1}M^3}$,
 for which ({\ref{X}}) gives
\beq
\dot{A}=2a\tanh(ay)\left\{-2+\frac{1}{3}(8-\xi^{-1})\cosh^{-2/3}(ay){\,}_2F_1(1/2,\,1/3,\,3/2,\,\tanh^2(ay))\,\right\}\,.
\eeq
Integrating, we obtain
\beq
A(y)=-4\ln(\cosh(ay))+\frac{1}{3}(8-\xi^{-1})\tanh^2(ay)\,F_{PFQ}\left(\{1,1,7/6\},\,\{3/2,2\},\,\tanh^2(ay)\right),
\eeq
The warp factor $e^{A}$ is plotted in Figure 7 for $\xi=1/4$  and for $\xi=1/9$.

 \begin{figure}[!h]
\centering
   \begin{minipage}[c]{0.5\textwidth}
   \centering \includegraphics[width=1\textwidth]{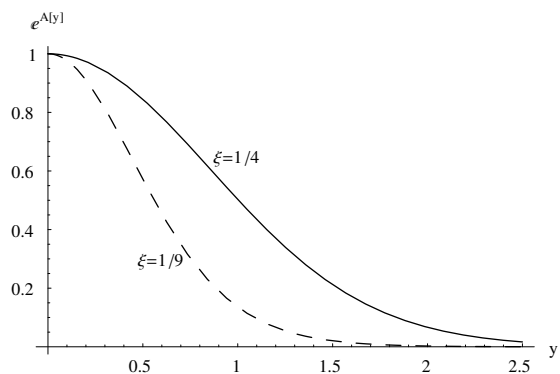}
   \end{minipage}%
   \caption{\textbf{Warp factors for $\bf{\phi( y ) = \phi( 0)\left(\cosh ( ay)\right)^{-1}}$,
   $\xi<1/2$.} }
   \label{figure7}
\end{figure}%
\newpage

For larger $\xi$ the behaviour does not change drastically. Note though that for values $\xi>1/2$ the warp
factor develops a maximum beyond the origin. In Figure 8 we plot the cases $\xi=1$ and $\xi=\infty$

 \begin{figure}[!h]
\centering
   \begin{minipage}[c]{0.5\textwidth}
   \centering \includegraphics[width=1\textwidth]{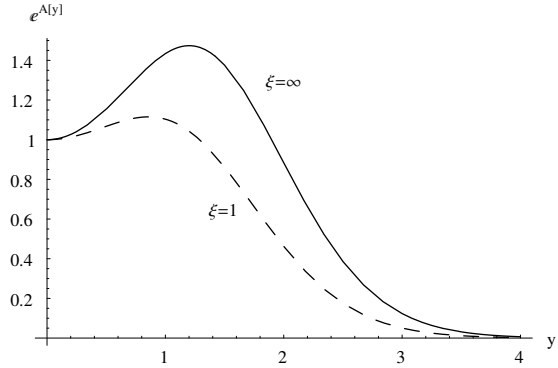}
   \end{minipage}%

   \caption{\textbf{Warp factors for $\bf{\phi( y ) = \phi( 0 )\left(\cosh ( ay)\right)^{-1}}$, $\xi>1/2$. }}
   \label{figure8}
\end{figure}%

As already seen, the equation of motion for a kink-like scalar cannot be solved analytically for general boundary conditions. 
It is however possible to obtain numerical solutions. We expect to find a set of ${\cal{Z}}_2$-symmetric solutions
for $e^{A( y)}$ that reduce to the known $\xi=0$ solution mentioned in \cite{KT}. 
As an example, we consider numerical solutions of the warp factor equation, imposing the boundary 
values $A(0)=1$, $A'(0)=0$ and taking different values for $\phi_0$. The resulting warp factors
for $\xi=-2$ and $\xi=0.8$ are shown in Figure 9. We have taken $\phi_0=a=1$ in units of $2M^3$.
 For this choice of boundary values and units, $\xi=2$ corresponds to the limiting value for
which the function $f(\phi)$ becomes zero at the origin, so higher values of $\xi$ are forbidden.
 Notice the peak beyond the origin in the second plot.

  \begin{figure}[!h]
\centering
   \begin{minipage}[c]{0.5\textwidth}
   \centering \includegraphics[width=1\textwidth]{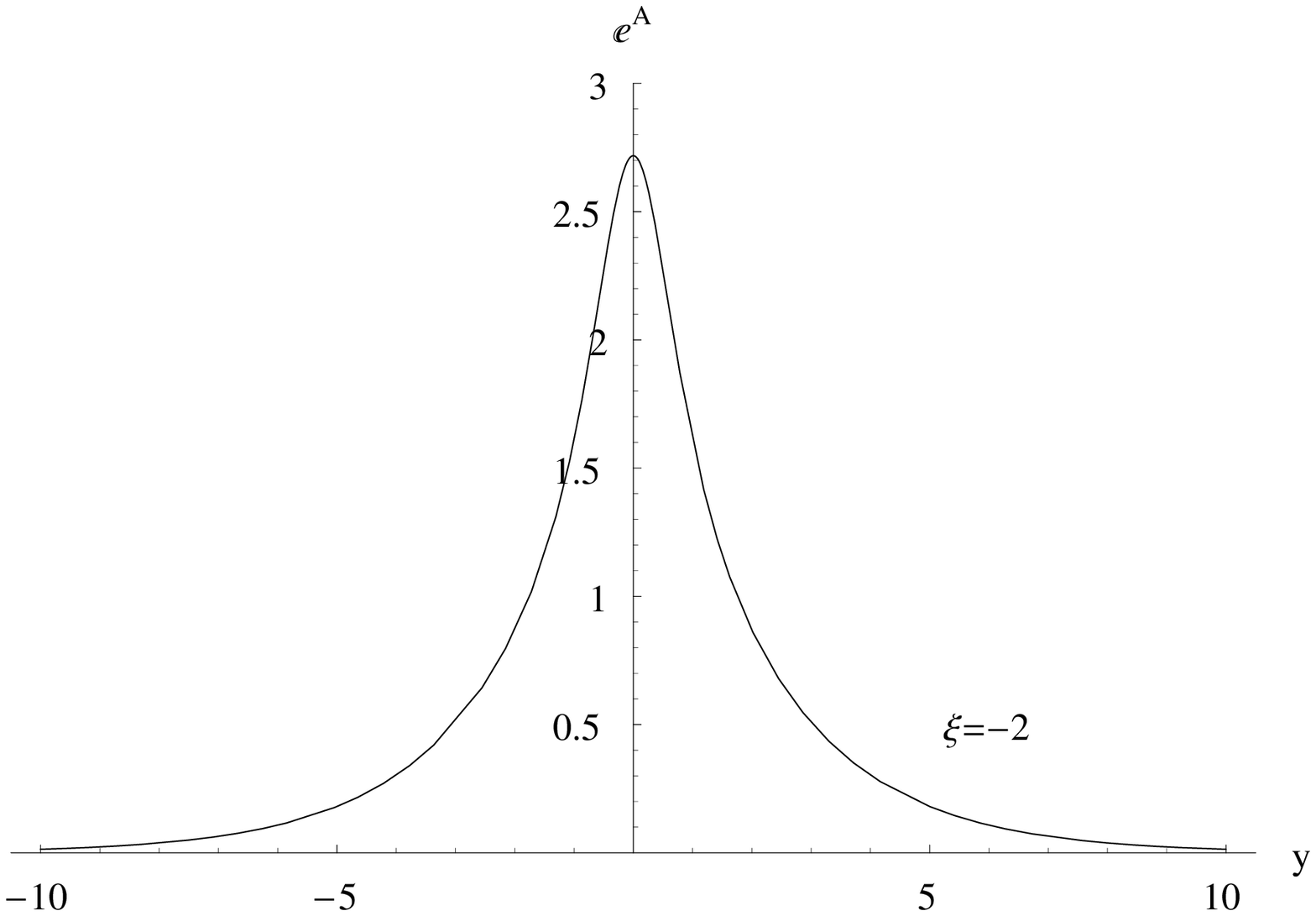}
   \end{minipage}%
      \begin{minipage}[c]{0.5\textwidth}
   \centering \includegraphics[width=1\textwidth]{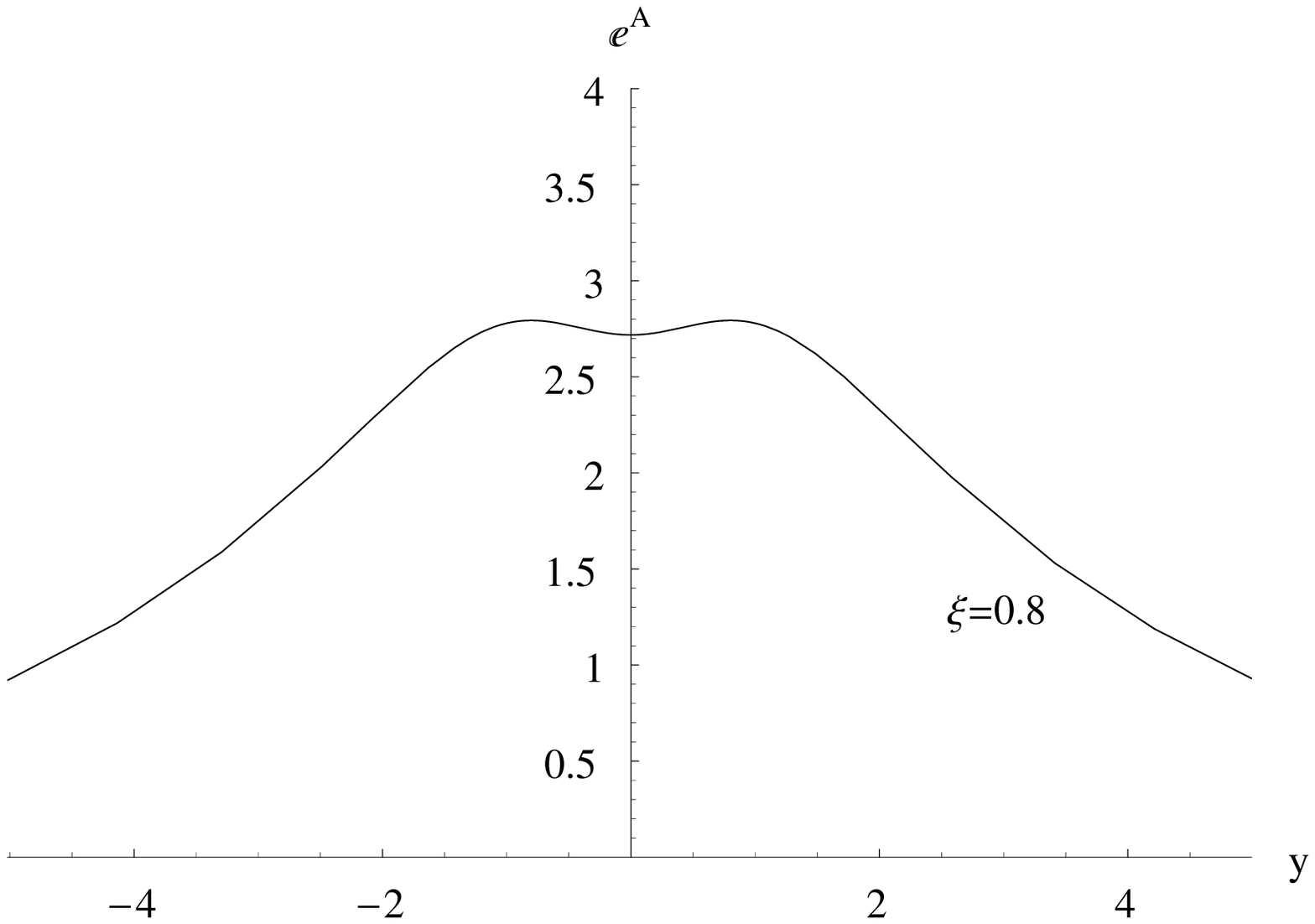}
   \end{minipage}%

   \caption{\textbf{Smooth numerical solutions.}}
   \label{figure9}
\end{figure}%

\section{Graviton Localization}

It would be interesting to check if gravity is localized in the geometries we calculated and especially 
those that deviate from the 
original Randall-Sundrum. Let us quickly provide a general argument and then study some specific examples.
 We consider a perturbation

\beq
\delta G_{{\rm M}{\rm N}}  = \delta ^\mu  _M \delta ^\nu  _N h_{\mu \nu } \left( {x,y} \right),
 \,\,\,\,\,\,\,\,\,\,   \delta \phi  = 0\,,
\eeq
for the gauge $h_{5M}=0$. Imposing transversality ( $h^\mu  _\mu   = \partial _\mu  h^{\mu \nu }  = 0$ ), we obtain to
first order, 
\beq
\left( { - \frac{{d^2 }}
{{dy^2 }} - e^{ - A\left( y \right)} \partial ^2  + \ddot A\left( y \right) +
 \dot A^2 \left( y \right)} \right)h_{\mu \nu }  = 0\,,
\eeq
where $\partial ^2  = \eta _{\mu \nu } \partial ^\mu  \partial ^\nu  $. 
Notice that this result is independent of the coupling
function $f(\phi))$. If we introduce a trial solution of the form of a product 
$h_{\mu \nu }  \propto e^{ip\cdot x}  \psi(y)$, we get a Schroedinger-like equation
\beq
\left( { - \frac{{d^2 }}
{{dy^2 }} + \ddot A(y) + {\dot{A}}^2(y)} \right)\psi(y)  =
 m^2 e^{ - A(y)} \psi(y)\,,
\eeq
where we have introduced the mass $m^{2}=-p^{2}$. In order to study the spectrum 
of this equation, it is more convenient to transform it into a conventional
 Schroedinger equation. In order to eliminate the exponential, we may introduce the transformation
 \beq
 \frac{d}{dy}=e^{-A/2}\frac{d}{dz}\,\,,\,\,\,\,\,\,\,\psi(y)=e^{\,A/4}\,\overline{\psi}\,.
 \eeq
The
resulting equation is 
\beq
\left(-\frac{d^2}{dz^2}+U(z)\,\right)\overline{\psi}=m^2\overline{\psi}\,,
\eeq
with the potential
\beq
U(z)=\frac{3}{4}\frac{d^2A}{dz^2}+\frac{9}{16}\left(\frac{dA}{dz}\right)^2\,.
\eeq	
Note that this equation can be put into the form
\beq
\left(-\frac{d}{dz}-\frac{3}{4}\frac{dA}{dz}\right)
\left(\,\frac{d}{dz}-\frac{3}{4}\frac{dA}{dz}\right)\,\overline{\psi}=m^2\overline{\psi}\,.
\eeq
This is supersymmetric Quantum Mechanics and the transformed graviton wavefunction (zero mode) corresponds to the supersymmetric ground state.
This form also excludes the existence of tachyon modes. The zero mode is just $\psi _0(y)= Ne^{A(y)}$ and it is normalizable.
  We also have to know if there is a gap between the zero mode and the continuum of eigenstates. 
  For this, we have to know the behaviour of the
  potential $U(z)$. Although for most of the cases above the change
   of variable $z=\int\,dy\,e^{-A(y)/2}$ is not analytically integrable, we may draw some conclusions with the help of
   the asymptotic behaviour ({\ref{ASYMPT}}). Since, for $y\rightarrow \infty$, we may have 
   \beq
   z=\int\,dy\,e^{-A/2}\approx
   \int\,dy\,e^{\overline{\kappa}y/2}\,(g(y))^{-1/2}
\eeq
or $z\propto \,e^{\overline{\kappa}y/2}$. As a result, 
\beq
\lim_{y\rightarrow\infty}U(z)\,\propto\,e^{A}\,\rightarrow \,0\,.
\eeq
Therefore, the continuous spectrum starts from zero mass and there is no gap.

  Next, we may check the profile of the localization potential $U(z)$ for various values of $\xi$. For $\xi=\frac{1}{8}$, we have
  $A(y)=-4ln(cosh(ay))$. In this case, the integration can be done analytically and the transformed coordinate is $z = \frac{1}{{4a}}\sinh 
  \left( {2ay} \right) + \frac{y}{2}$. For $\xi=0$, the warp function becomes $A\left( y \right) =  - \frac{4}{9}\left( {4\ln \left( {\cosh \left( {ay} \right)} \right) +
   \tanh ^2 \left( {ay} \right)} \right)$. We can only proceed numerically to perform the change of coordinates
and calculate the potential. The resulting profiles are depicted in Figure 10. The localizing potential has the familiar volcano-like
shape we also encounter in standard Randall-Sundrum.

  \begin{figure}[!h]
\centering
   \begin{minipage}[c]{0.5\textwidth}
   \centering \includegraphics[width=1\textwidth]{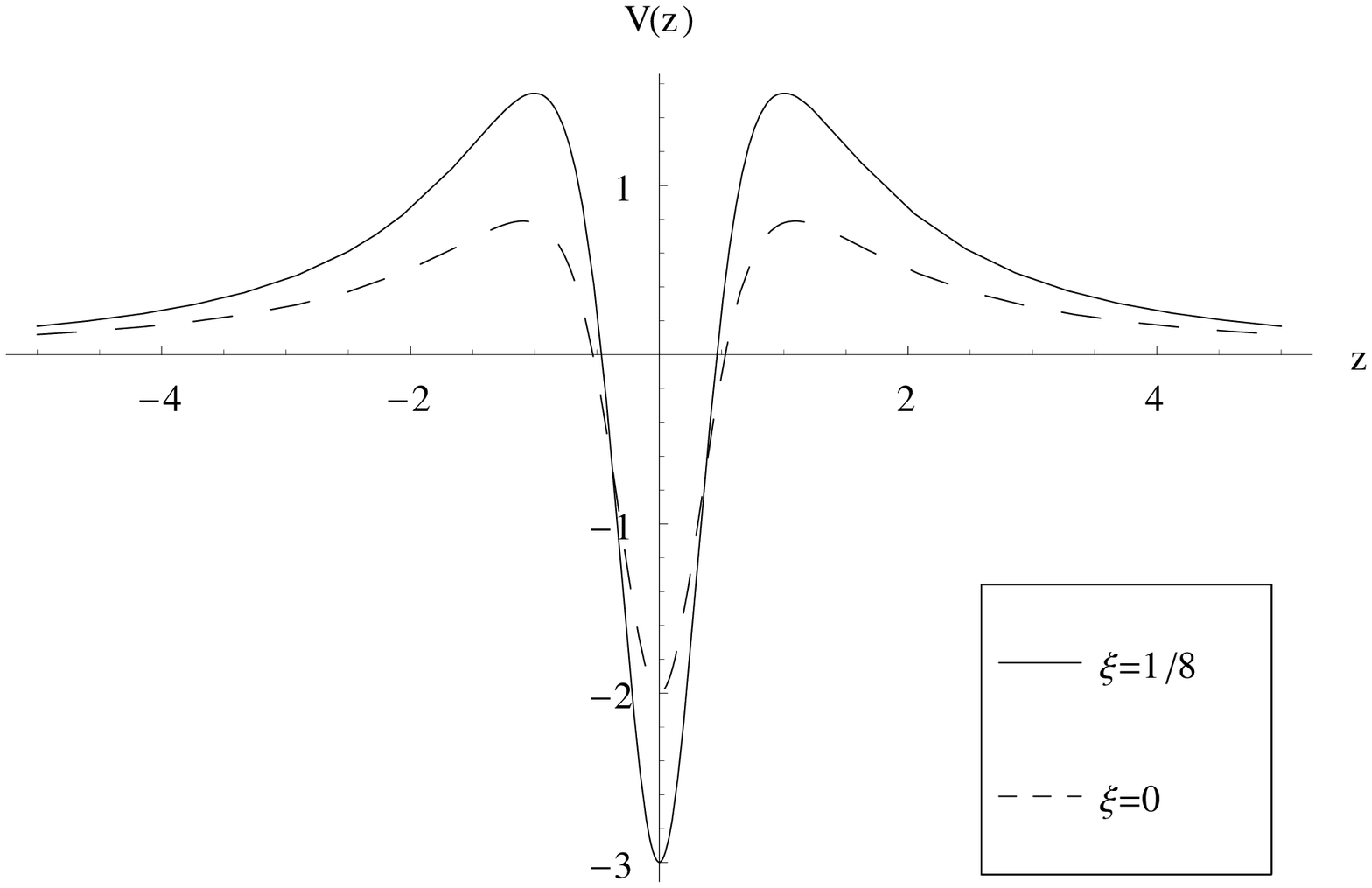}
   \end{minipage}%
      \begin{minipage}[c]{0.5\textwidth}
   \centering \includegraphics[width=1\textwidth]{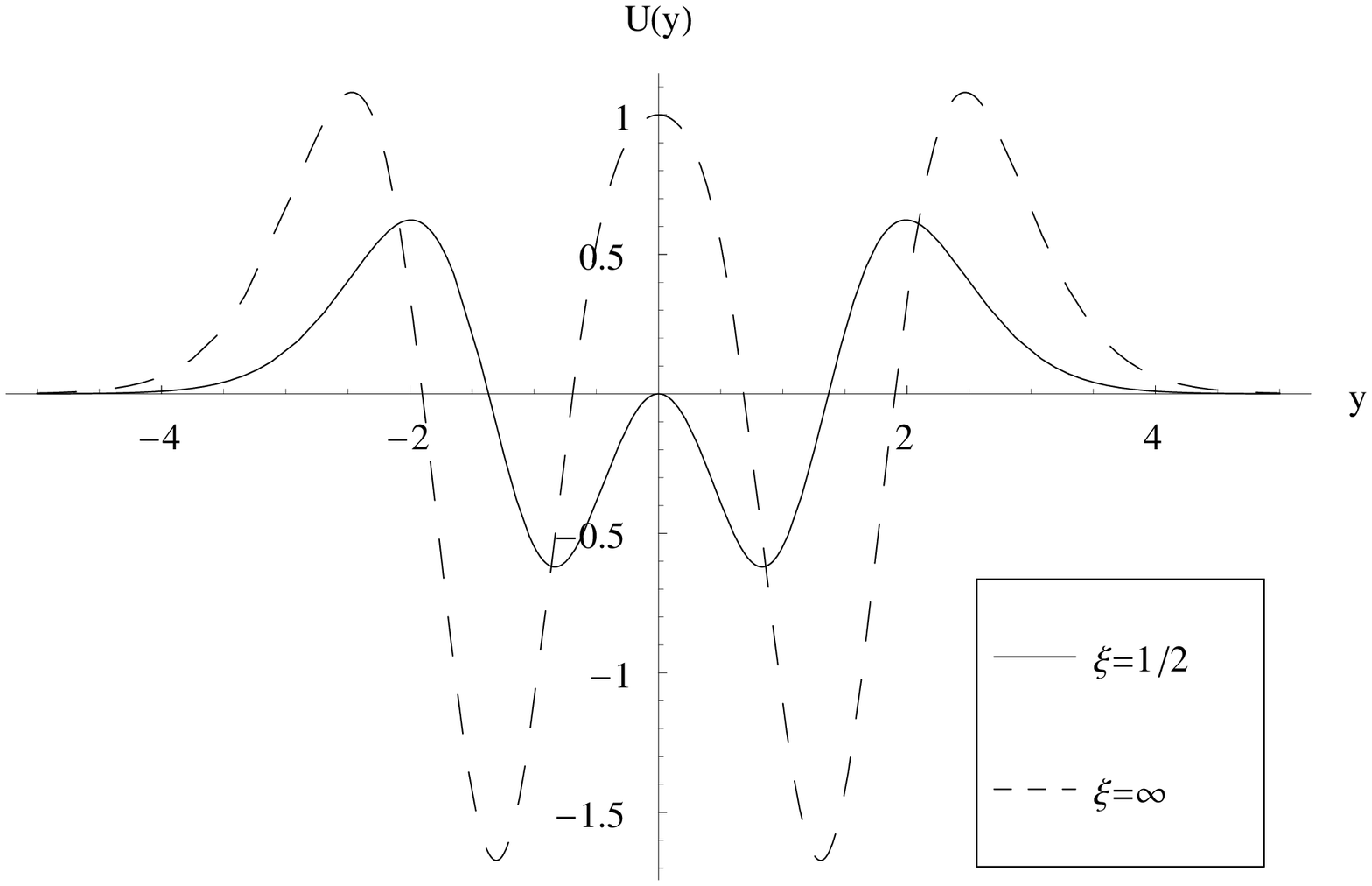}
   \end{minipage}%

   \caption{\textbf{Localizing potentials for various values of $\xi$.}}
   \label{figure10}
\end{figure}%

  It turns out that the volcano-like profile of the potential is not maintained for all values of $\xi$. 
  For the boundary value and unit choices made for the smooth numerical solutions of section 5 depicted in Figure 9, 
  we find that at the value $\xi=(\sqrt {193}- 9)/16\approx 0.306$
  the global minimum at the origin $y=0$ changes into a local maximum. Thus, as we move towards higher $\xi$'s,
   a central spike is developed. For $\xi=1/2$ the potential becomes zero on the {\textit{Brane}}, 
   while as $\xi$ goes to infinity the potential at that point approaches unity. The
   corresponding graphs are shown in Figure 10.

\section{Conclusions}

 In the present article we investigated the existence of solutions for a non-minimally coupled Bulk scalar field 
 in a warped Brane-world framework. For a scalar field coupling to gravity of the form $-\frac{1}{2}\xi \phi^{2}R$, we 
 derived a set of non-singular solutions for a wide range of the coupling parameter $\xi$. 
 We demonstrated the compatibility of the usual Randall-Sundrum
warp factor with the presence of a non-trivial scalar field for a suitably chosen scalar potential. This was done for
either  a scalar field-dependent or independent Brane-tension. 
The profile of the scalar field solution in the field-independent Brane-tension case is that of a
 {\textit{folded-kink}} ( $\tanh(a|y|)$ ). The scalar field 
acquires its minimum value on the brane, approaching a constant value as we move towards infinity in the y-direction.
 The {\textit{conformal value}} of the coupling parameter $\xi_{c}$ separates the above mentioned solutions from singular
 {\textit{``solutions"}}.  Thus, for the Randall-Sundrum warp factor, non-singular scalar field solutions exist only for 
 $\xi>\xi_c$ and they are, in general, of the above folded-kink shape. 
 For negative values of the coupling parameter $\xi$
the scalar field {\textit{``solutions"}} found correspond to a scalar potential with negative powers of the scalar field
and, therefore, singular. A field-dependent Brane-tension allows 
for a more diverse range of behaviour including scalar field solutions which exponentially decrease at infinity.

  Guided by the scalar field set of solutions with a folded kink type of profile, we investigated the existence of
  general warp factor solutions that are different from the exact Randall-Sundrum case but still localized. Thus, 
  assuming a scalar field solution of the form $\tanh(a|y|)$, we derived corresponding warp factor solutions which we
  analyzed semi-analytically and numerically. Our analytical treatment further showed that for a wide range of values in the parameter space of the model we get finite 
   geometries which are well-behaved
for large y, as long as the Brane tension we introduce is large enough. Furthermore, we considered smooth warp factor solutions in which the role of 
  the Brane is played by the scalar field itself. In this setup we considered some special solutions and proceeded to study
  numerically general finite geometries. We concentrated on ${\cal{Z}}_2$-symmetric solutions, although asymmetric ones are
  also possible. We considered a class of solutions which asymptotically reduce to decreasing exponentials of the
   Randall-Sundrum type. These solutions exist for a coupling parameter $\xi$ within a range of values.  For a subset of
   these localized solutions the warp factor is not a monotonous decreasing function but exhibits a second maximum close to
   but beyond the origin and subsequently decreases. We have also derived analytically special exact solutions existing 
   for special choices of boundary values and for a range of the coupling parameter. For these solutions, the same warp
   factor corresponds to either a kink scalar solution or the solution $\phi=\phi(0)\left(\cosh(ay)\right)^{-1}$.
   
  Finally we considered the localization of gravitons near the brane. Although, the Schroedinger-like equation 
  for gravitational perturbations is the same as in the minimal case, the warp factor detailed profile depends on the
  coupling parameter and the details of the localized spectrum should depend on it. Of course, again 
  the spectrum has no mass gap and does not contain any tachyonic modes. The form of the localizing
(volcano) potential depends on the detailed profile of the warp factor. It was studied numerically 
in a number of cases but also analytically in special cases. 
For a particular choice of boundary scalar field value and the special coupling parameter value 
$\xi=\frac{1}{8}$, the localizing potential has the typical $\xi=0$ volcano profile. Nevertheless,
 for values of $\xi$ larger than a certain value, the localizing potential develops
  a spike at the origin, that increases along with $\xi$. 
  For $\xi=\frac{1}{2}$ the spike reaches zero, while it tends to one for very large values of $\xi$. This behaviour is currently
  being studied and will be the subject of a future publication.

\bigskip

{\bf Acknowledgments.} This research was co-funded by the European Union
in the framework of the Program $\Pi Y\Theta A\Gamma O PA\Sigma-II$ 
of the {\textit{``Operational Program for Education and Initial
Vocational Training"}} ($E\Pi EAEK$) of the 3rd Community Support Framework
of the Hellenic Ministry of Education, funded by $25\%$ from 
national sources and by $75\%$ from the European Social Fund (ESF). C. B.
aknowledges also an {\textit{Onassis Foundation}} fellowship.

\end{document}